\documentclass[aps,prl,twocolumn,a4paper,10pt,notitlepage,footinbib,superscriptaddress,showpacs]{revtex4-1}%
\usepackage[english]{babel}
\usepackage[T1]{fontenc}
\usepackage{endnotes}
\usepackage{amssymb,amsmath,amsfonts}
\usepackage{textcomp}
\usepackage{graphicx,color}
\usepackage[utf8x]{inputenc}
\usepackage{float}
\usepackage{subfigure}

\begin{document}

\title{Lamellar ordering, droplet formation and phase inversion in exotic active emulsions}

\author{Francesco Bonelli}
\affiliation{Dipartimento di Meccanica, Matematica e Management, DMMM, Politecnico di Bari, 70125 Bari, Italy}

\author{Livio Nicola Carenza}
\affiliation{Dipartimento di Fisica and Sezione INFN, Università degli Studi di Bari, 70126 Bari, Italy}

\author{Giuseppe Gonnella}
\affiliation{Dipartimento di Fisica and Sezione INFN, Università degli Studi di Bari, 70126 Bari, Italy}

\author{Davide Marenduzzo}
\affiliation{SUPA, School of Physics and Astronomy, University of Edinburgh, Edinburgh EH9 3FD, United Kingdom}

\author{Enzo Orlandini}

\author{Adriano Tiribocchi}
\affiliation{Dipartimento di Fisica and Sezione INFN, Università degli Studi di Padova, I-35131 Padova, Italy}

\date{\today}

\begin{abstract}
We study numerically the behaviour of a mixture of a passive isotropic fluid and an active polar gel, in the presence of a surfactant favouring emulsification. Focussing on parameters for which the underlying free energy favours the lamellar phase in the passive limit, we show that the interplay between nonequilibrium and thermodynamic forces creates a range of multifarious exotic emulsions. When the active component is contractile (e.g., an actomyosin solution), moderate activity enhances the efficiency of lamellar ordering, whereas strong activity favours the creation of passive droplets within an active matrix. For extensile activity (occurring, e.g., in microtubule-motor suspensions), instead, we observe an emulsion of spontaneously rotating droplets of different size. By tuning the overall composition, we can create high internal phase emulsions, which undergo sudden phase inversion when activity is switched off. Therefore, we find that activity provides a single control parameter to design composite materials with a strikingly rich range of morphologies.
\end{abstract}

\maketitle

Active matter has established itself as a topical area of research in physics over the last few years~\cite{RMP}. Active systems are internally driven, and continuously take up energy from their surrounding so that they function far from thermal equilibrium. Examples are bacterial swarms~\cite{Cisneros}, cell extracts~\cite{Surrey}, cytoskeletal gels~\cite{Dogic,Ignes} and chemically driven phoretic colloids~\cite{synthetics}. Their inherent non-equilibrium nature causes a range of unexpected behaviours, such as spontaneous flows~\cite{Spfl}, bacterial turbulence~\cite{Cisneros,turbulence} and motility-induced phase separation~\cite{APS}.

While single-component active systems have received a lot of attention, much less is currently known about the behaviour of mixtures, made up by a combination of active and passive components. Mixtures of self-propelled and passive colloidal spheres and rods have been studied via particle-based simulations in~\cite{baskaran,joakim,joanny}, finding that activity can drive self-assembly or may trigger segregation between the passive and active components. Lyotropic binary mixtures with an active components have also been considered in Refs.~\cite{elsen,adriano,matthew,giomi} within a continuum model, showing that activity creates self-motile droplets, and may also cause an undulatory instability of the passive-active interface. In general, these composite materials bear great promise as self-assembling new soft materials, as activity allows the system to bypass thermodynamic constraints which would otherwise govern its behaviour, and leads to novel phenomena.

Here we study a theoretical model for a new kind of active material, made up by mixing an isotropic passive fluid with a polar active one, in the presence of a surfactant which favours emulsification. The model is based on a generalization of the Brazovskii free-energy functional~\cite{braz}  useful  to describe  complex fluids where the presence of interfaces is favored. We name the resulting composite material an ``exotic active emulsion''. There are two potential avenues to realise these systems in the lab. The first is by dispersing sticky bacteria~\cite{jana} or self-attractive cytoskeletal gels~\cite{Dogic,Ignes} (the active fluid) in water (the isotropic component), under conditions promoting microphase separation (e.g., depletion forces~\cite{Dogic,Ignes,jana} and a suitable surfactant).
The second route is through activation of passive emulsions: for instance, it is now possible to encapsulate an active nematic gel within a water-in-oil emulsion~\cite{Dogic,Ignes}. We shall comment more on potential experiments at the end of our work.

Active fluids can be either contractile (actomyosin gels~\cite{gowrishankar}) or extensile (bacteria~\cite{jana} or microtubule-kinesin~\cite{Dogic,Ignes} suspensions), according to the nature of the internal driving force~\cite{RMP}. By systematically varying the strength and nature of activity and the mixture composition, we show here that these two situations lead to qualitatively different instances of exotic active emulsions, with a wide range of surprising and unexpected behaviours. Starting from a lamellar phase (with a 50:50 ratio between the active and passive components), dialling up contractile activity first enhances lamellar ordering, and then creates an emulsion of passive droplets embedded in an active, spontaneously flowing, background. Extensile activity, instead, creates an emulsion of active rotating droplets in a passive matrix. In this case, by varying the mixture composition, we can additionally create ``high-internal phase'' emulsions, where the active component is dispersed in droplets even when it constitutes the majority phase. All these morphological changes correspond to nonequilibrium phase transitions, whilst the emerging domain size in each phase is mainly controlled by activity.

The physics of an exotic polar active emulsion can be described by using an extended version of the well-established active gel theory~\cite{RMP,elsen,adriano}, in which a set of balance equations governs the evolution of the following hydrodynamic variables: the density of the fluid $\rho({\bf r},t)$ and its velocity ${\bf v}({\bf r},t)$, the  concentration of the active material $\phi({\bf r},t)$ and the polarization ${\bf P}({\bf r},t)$, which determines its average orientation. The equilibrium properties of the system are encoded in the following Landau-Brazovskii~\cite{prl_yeom,braz} free-energy functional
\begin{eqnarray}\label{fe}
&F&[\phi,\mathbf{P}]
=\int d^{3}r\,\{\frac{a}{4\phi_{cr}^4}\phi^{2}(\phi-\phi_0)^2+\frac{k}{2}\left|\nabla \phi\right|^{2}+\frac{c}{2}(\nabla^2\phi)^2 \nonumber\\
&-&\frac{\alpha}{2} \frac{(\phi-\phi_{cr})}{\phi_{cr}}\left|\mathbf{P}\right|^2+ \frac{\alpha}{4}\left|\mathbf{P}\right|^{4}+\frac{\kappa}{2}(\nabla\mathbf{P})^{2}
+\beta\mathbf{P}\cdot\nabla\phi\},
\end{eqnarray}
where the first term, multiplied by the phenomenological constant $a>0$, describes the bulk properties of the lamellar fluid while the second and third terms determine the interfacial tension. Notice that here $k<0$ in order to favour the formation of interfaces while a positive value of $c$ guarantees the stability of the free-energy \cite{braz}. The bulk term is chosen in order to create two free energy minima,  $\phi=0$ and $\phi=\phi_0$ corresponding respectively to the passive and active phase. Note that $\phi_{cr}=\phi_0/2$, where $\phi_{cr}$ is the critical concentration for the transition from isotropic ($|{\bf P}|=0$) to polar ($|{\bf P}|>0$) states.
 The bulk properties of the polar liquid crystal are instead controlled by the $|\textbf{P}|^2$ and $|\mathbf{P}|^{4}$ terms, multiplied by the positive constant $\alpha$, whereas the energetic cost due to the elasticity is gauged by the constant $\kappa$ (in the single elastic constant approximation~\cite{deGennes}). Finally the last term takes into account the orientation of the polarization at the interface of the fluid. If $\beta\ne 0$, $\textbf{P}$ preferentially points perpendicularly to the interface (normal anchoring): towards the passive (active) phase if $\beta>0$ ($\beta<0$). 

\begin{figure}[b!]
\centering
\includegraphics[width = 0.5\textwidth]{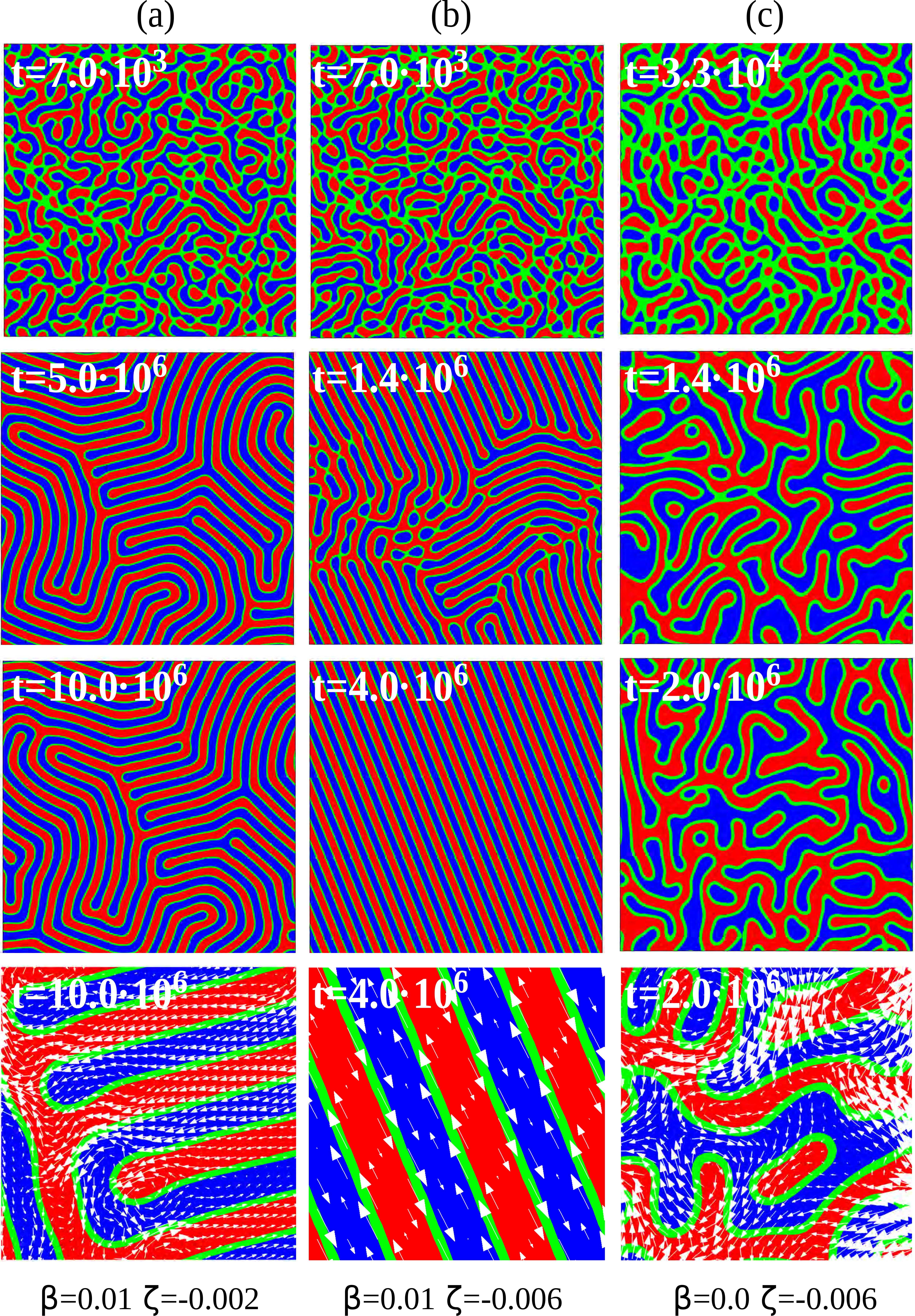}
\caption{Contour plots of $\phi$ at various times, for 50:50 composition mixtures with small contractile activity. Each column represents a different physical situation: (a) and (b) show a contractile emulsion with normal anchoring and a different value of $\zeta$; (c) shows a contractile emulsion with no anchoring. In (b) normal anchoring and moderate activity favour the formation of a regular array of lamellae. The last row shows a zoom of the velocity field in the system. These simulations have been performed on a square lattice of size $L=128$. The color range of the contour plot is the same for all figures and is defined as follows: red (active phase) if $\phi>1.5$, green if $0.5<\phi<1.5$, blue (passive phase) if $\phi<0.5$.}
\label{fig:fig.1}
\end{figure}

\begin{figure}[htbp]
\centering
\includegraphics[width=1.0\linewidth]{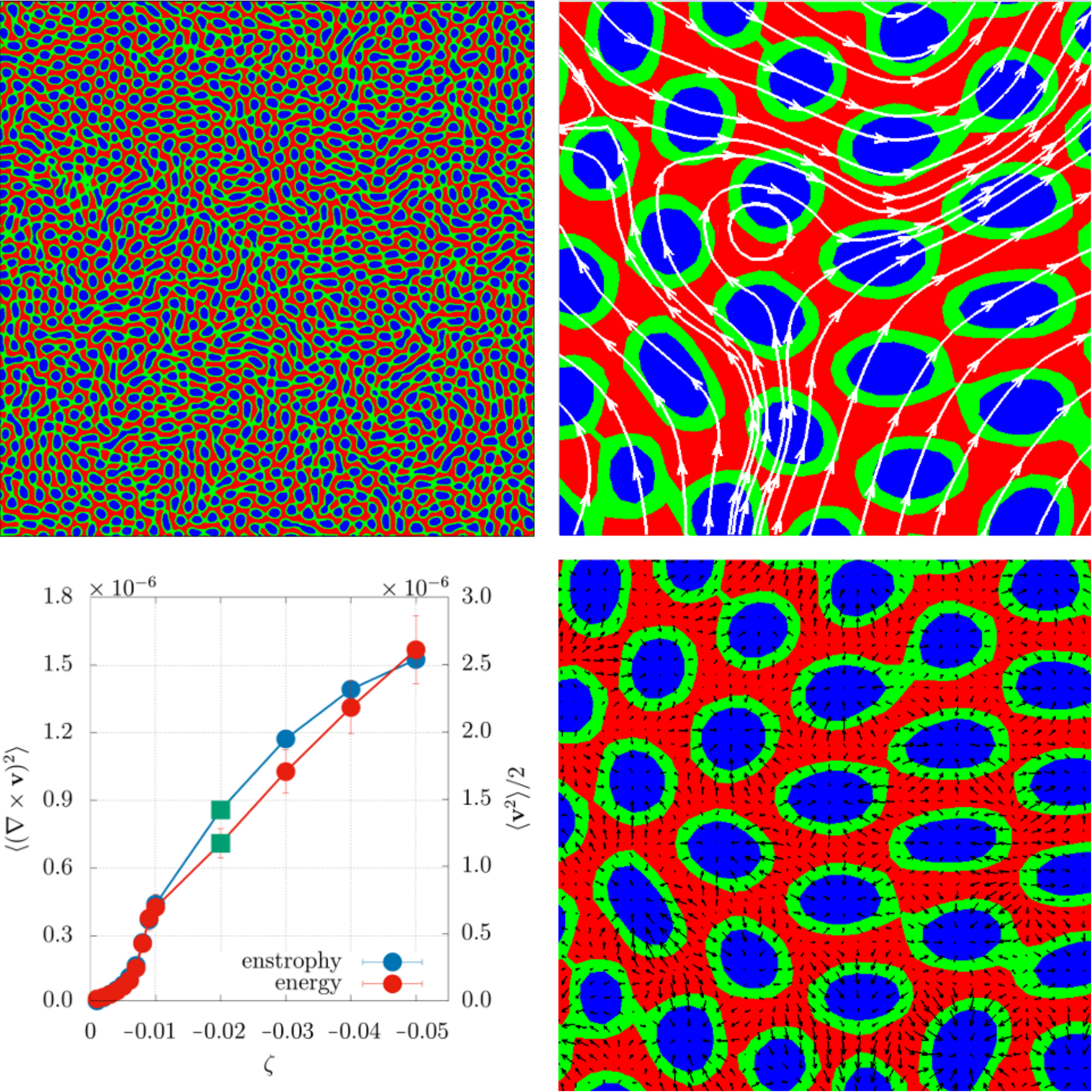}
\caption{(Top, left) Late time contour plot of $\phi$, for a $50:50$ mixture with strong contractile activity ($\beta=0.01$, $\zeta=-0.02$), showing a self-assembled emulsion of passive droplets in an active matrix. (Top, right)  Zoom with streamlines of the flow field. (Bottom, left) Plot of the mean kinetic energy and of the mean enstrophy versus $\zeta$. (Bottom, right) Zoom of the system with the $\textbf{P}$ field. Simulations were performed on a square lattice of size $L=256$.}
\label{fig:fig.2}
\end{figure}

\begin{figure}[htbp]
\centering
\includegraphics[width=1.0\linewidth]{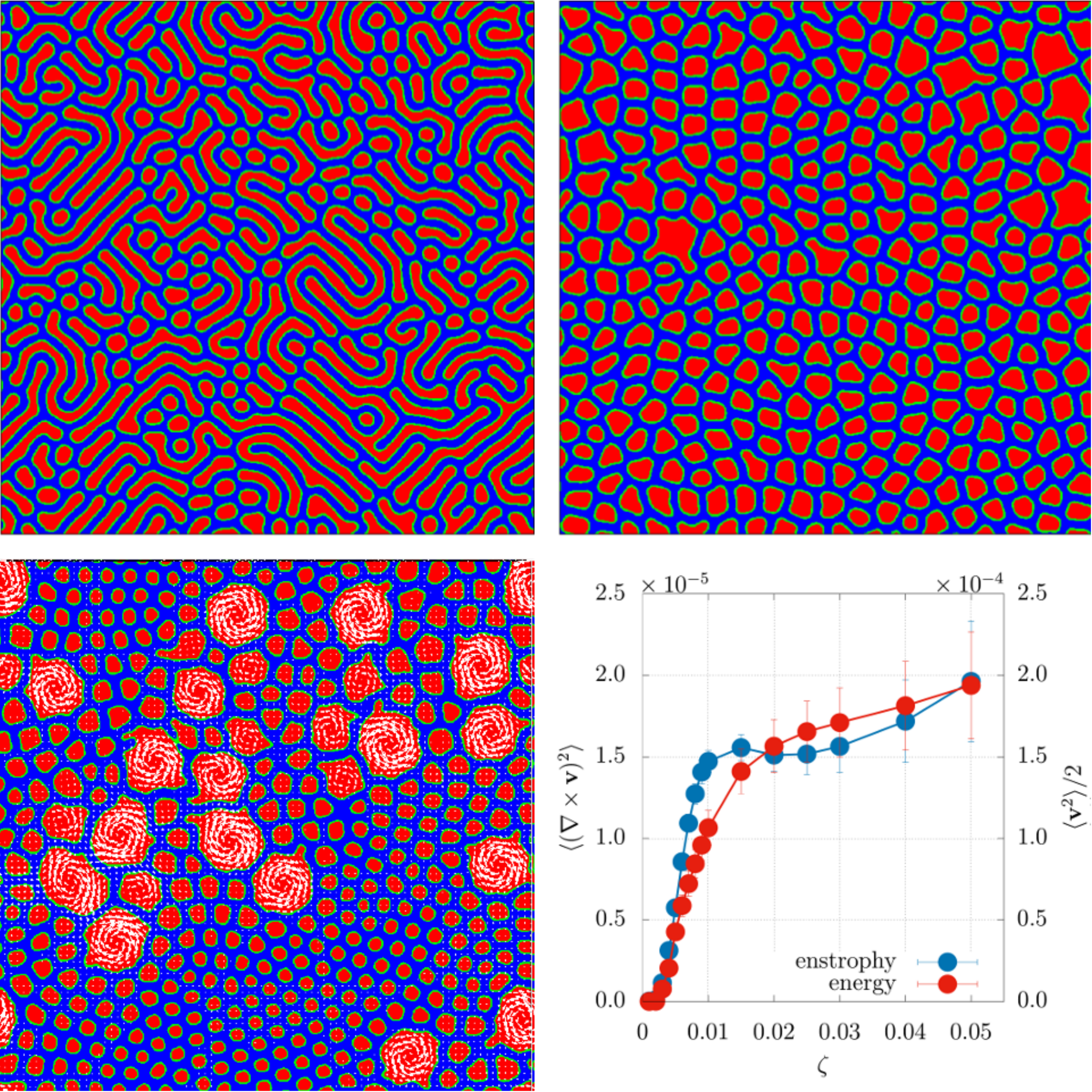}
\caption{Late time contour plots of $\phi$ for $50:50$ extensile mixtures. (Top, left) $\beta=0.01$ $\zeta=0.001$; (top, right) $\beta=0.01$, $\zeta=0.002$; (bottom, left) $\beta=0.01$, $\zeta=0.003$. Extensile mixture lead to an emulsion of active droplets within a passive background. For moderate activity the larger droplets rotate, here in an anti-clockwise sense. 
Arrows indicate the direction of the velocity field. (Bottom, right) Plot of the mean kinetic energy and of the mean enstrophy for several values of the extensile activity. Simulations were performed on a square lattice of size $L=256$.}
\label{fig:fig.3}
\end{figure}

The dynamic equations governing the physics of the system are
\begin{eqnarray}
\rho\left(\frac{\partial}{\partial t}+\mathbf{v}\cdot\nabla\right)\mathbf{v} & = & -\nabla P+\nabla\cdot\underline{\underline{\sigma}}^{total}.\label{nav}\\
\frac{\partial \phi}{\partial t}+\nabla\cdot\left(\phi\mathbf{v}\right)&=&\nabla\left( M\nabla\frac{\delta F}{\delta \phi}\right),\label{conc_eq}\\
\frac{\partial\mathbf{P}}{\partial t}+\left(\mathbf{v}\cdot\nabla\right)\mathbf{P}&=&-\underline{\underline{\Omega}}\cdot\mathbf{P}+\xi\underline{\underline{D}}\cdot\mathbf{P}
-\frac{1}{\Gamma}\frac{\delta F}{\delta\mathbf{P}},\label{P_eq}
\end{eqnarray}
in the limit of incompressible fluid. The first one is the Navier-Stokes equation, in which $P$ is the ideal gas pressure and $\underline{\underline{\sigma}}^{total}$ is the total stress tensor~\cite{supplement}. The latter includes an active contribution $-\zeta {\mathbf P}{\mathbf P}$, where $\zeta$ is the activity, which is positive (negative) for extensile (contractile) fluids -- this plays a key role in determining the system behaviour.
Eqs.~\ref{conc_eq}-\ref{P_eq} govern the evolution of the concentration of the active material and of the polarization field, respectively.
The former is a convection-diffusion equation in which $M$ is the mobility and $\mu=\delta F/\delta\phi$ is the chemical potential. The latter is an advection-relaxation equation where $\Gamma$ is the rotational viscosity, {$\xi$ is a constant controlling the aspect ratio of active particles (positive for rod-like particles and negative for disk-like ones)},
$\textbf{h}=\delta F/\delta\textbf{P}$ is the molecular field~\cite{deGennes}, and   $\underline{\underline{D}}=(\underline{\underline{W}}+\underline{\underline{W}}^T)/2$ and  
$\underline{\underline{\Omega}}=(\underline{\underline{W}}-\underline{\underline{W}}^T)/2$ represent the symmetric and the antisymmetric part of the velocity gradient tensor 
$\underline{\underline{W}}=\nabla{\bf v}$. These contributions are in addition to the material derivative as the liquid crystal can be rotated or aligned by the fluid~\cite{BerisEdwards}.

To begin with, we consider a symmetric (50:50) emulsion with weak contractile activity ($\zeta<0$). [All other parameters are given in~\cite{note} -- a mapping to physical units is given in~\cite{supplement}.] With normal anchoring (Fig.~1a,b,~Suppl.~Movies 1 and 2), and starting from a uniform phase with small random fluctuations, a disordered lamellar texture rapidly emerges at early times (Fig.~1a,b, top panels). Later on, the pathway followed is greatly affected by the magnitude of $|\zeta|$. If this is small, lamellar ordering arrests at late times: here active forces create vortex-like flows localised at grain boundaries (Fig.~1a, bottom panel). For moderate activity the structure orders into a regular, defect-free, array of lamellae with large spontaneous interfacial flow. Activity enhances lamellar ordering also for larger system size, although some defects survive until late times (see Fig.~S1~\cite{supplement}, where the structure factor is also shown).

Lamellar ordering is also affected by the strength of the anchoring. If this is absent, lamellae get disrupted, and domains either partially coarsen, or break up into droplets (Fig.~1c, Suppl.~Movie 3).
At steady state, the mixture is morphologically closer to a bicontinuous phase, reminiscent of arrested spinodal decomposition. The active flow pattern is significantly different from those in Figures 1a,b: rather than being confined at the boundary between lamellae, the flow is now driven by deformations in the polarisation which tend to occur inside active domains. 

Systematic scanning of the activity for fixed normal anchoring (Figs.~2 and S2~\cite{supplement}) reveals the existence of another morphology for stronger activity, a self-assembled emulsion of passive droplets in an active matrix. The spontaneous flow within the active background keeps stirring the system (Figs.~2, top right, and~S4~\cite{supplement} for typical flow field plots), so that the passive droplets never settle into a static pattern. The flow profile is compatible with active turbulence~\cite{Thampi}. The transition between lamellar and droplet emulsions is signalled by a discontinuity in the plots of the mean enstrophy and kinetic energy as a function of $\zeta$ (at $\zeta\sim -0.007$ in Fig.~2, bottom left).

An analysis of the evolution of the morphology with time (Suppl.~Movie~4~\cite{supplement}), or of the steady state pattern with activity (Fig.~S2), suggests a mechanism for the transition between lamellar and droplet emulsion. If the activity is strong enough, undulations at the lamellar boundary favour interfacial splay, which in turn leads to active forces pinching off lamellae into droplets. [The dimensionless parameter controlling the pinch-off instability should therefore be $\zeta l/{\sigma}$, with $l$ the Brazovskii domain size and $\sigma$ the effective surface tension.] But why do passive, rather than active, droplets form? The reason is that splay creates inward asters for the polarization field: placing an isotropic droplet in the middle therefore relieves elastic stresses (Fig.~2, bottom right). The mechanism is therefore to some extent similar to the one which generically drives nanoparticles towards disclinations and defects in liquid crystals~\cite{bluephase}. Inspection of the structure factor of the droplet emulsions found at large $|\zeta|$ additionally suggests that the steady-state domain size decreases with contractile activity (Fig.~S3~\cite{supplement}).

We now consider the case in which the active component is extensile.
Switching the sign of the activity parameter creates completely different patterns. 
Upon increasing $\zeta$ the lamellar phase gives way to an emulsion of active droplets within a passive background (Figs.~3 and S5~\cite{supplement}) -- the inverse of what happens in contractile mixtures. The active droplet emulsion is approximately monodisperse for $\zeta=0.002$ (Fig.~3, top right), and bidisperse for $\zeta=0.003$ (Fig.~3, bottom left). The transition between the lamellar and emulsion morphologies corresponds to the value of activity for which the mean enstrophy and kinetic energy depart from zero, $\zeta_c\sim 0.002$ (Fig.~3). Analysis of the mixture structure factor further shows that the droplet size increases monotonically with $\zeta$ (Fig.~S6~\cite{supplement}) -- again the opposite behaviour of contractile mixtures. At very large $\zeta$ (as long as $\zeta\lesssim 0.02$) there is a crossover to macroscopic phase separation between active and passive components (corresponding to the plateau in the enstrophy and kinetic energy plots in Fig.~3). For higher values of $\zeta$  the demixing process stops.

The active component is now in droplets because extensile activity and normal anchoring lead to a purely rotational flow in the droplets, creating rotating spirals similar to those described in~\cite{kruse,kruse2} (see polarisation patterns in Fig.~S7 in\cite{supplement}).  Without this sustained flow, the internal spiral defect would be elastically unstable, as was the case for the contractile mixture (and this explains the inversion of the emulsion). The rotating droplet suspension self assembles spontaneously from an initially uniform mixture. In line with the theory of Ref.~\cite{kruse} we expect that a droplet of size $l$ should only rotate if $\zeta l^2/{\kappa}$ exceeds a critical threshold, whereas dimensional analysis suggests that the rotation velocity should be given by $\zeta/\eta$ -- both these predictions are borne out by an analysis of our numerical data (Figs.~S8~\cite{supplement}). Droplet shapes fluctuate greatly over time, and often fuse or split up during the dynamics (Suppl. Movies~5, 6\cite{supplement}). Thus, activity effectively decreases the interfacial surface tension.

\begin{figure}[htbp]
\centering
\includegraphics[width = 0.48\textwidth]{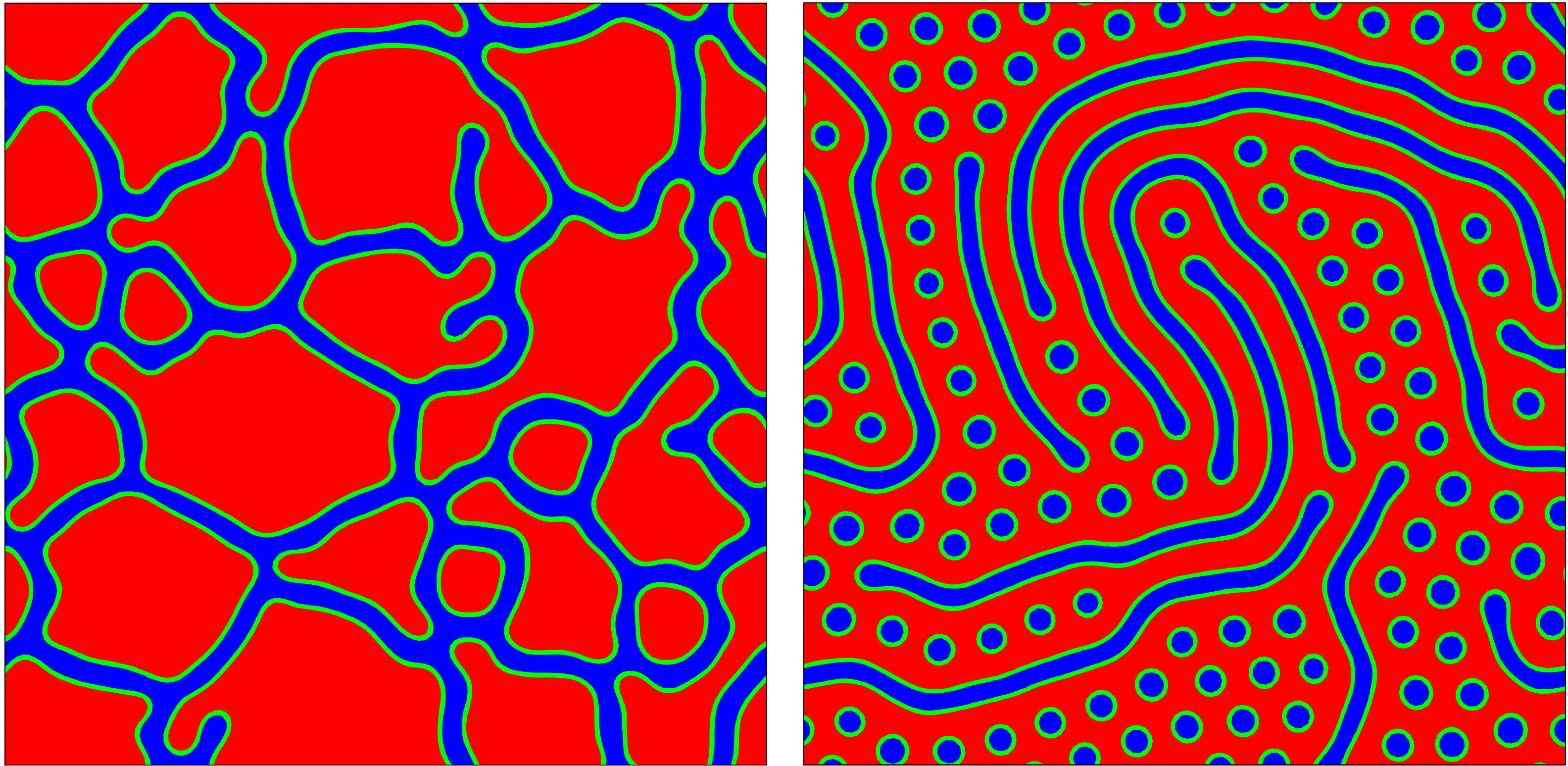}
\caption{Contour plots of $\phi$ for an 80:20 emulsion and $\beta=0.01$. (Left) Late time configuration for $\zeta=0.0015$. (Right) Configuration resulting from dynamic phase inversion. This is obtained starting from the configuration on the left, then  imposing a larger activity ($\zeta=0.008$), letting the system evolve (for $10^5$ timesteps), and finally switching the activity off ($\zeta=0$). Simulations have been perfomed on a square lattice of size $L=128$.}
\label{fig:fig.5}
\end{figure}

All results so far pertain to a 50:50 ratio between active and passive component. Changing the composition yields similar phenomenology. Focussing for concreteness on extensile mixtures, a 10:90 emulsion (where the first number refers to the active component) shows transitions from a lattice of non-motile droplets (the equilibrium configuration for a passive asymmetric mixture) to an active rotating droplet emulsion, with droplet size increasing with $\zeta$ as for 50:50 mixtures (Figs.~S9, S10~\cite{supplement}). Increasing the active component fraction beyond $50\%$ leads to additional interesting dynamics, as it creates high internal phase emulsions, i.e.  systems where the majority (active) phase forms droplets (Fig.~\ref{fig:fig.5}, for a 80:20 emulsion, and Fig.~S11~\cite{supplement} for a 70:30 emulsion). If the activity is first increased and then switched off, the emulsion undergoes a dynamic phase inversion, to leave a suspension of isotropic islands and droplets, within a liquid crystalline sea (see Fig.~\ref{fig:fig.5}). The pathway to phase inversion involves first rapid droplet coalescence, due to the turbulent spontaneous flow which arises at high activity~\cite{supplement}, followed by a slow reorganization driven by the chemical potential and thermodynamic stress tensor, when activity is switched off. This phenomenon is reminiscent of phase inversion in emulsions~\cite{salager}, although in our case this is triggered by variation of a nonequilibrium parameter (the activity).

In summary, here we have studied the behaviour of an emulsion obtained by mixing an isotropic fluid with an active polar gel, in the presence of a surfactant. The thermodynamic free energy density favours normal anchoring at the interface, as well as the formation of a lamellar phase, for symmetric passive mixtures. Our simulations predict that the resulting composite materials should display a striking variety of exotic phase behaviours and morphologies: remarkably, any one of these can be selected by tuning {\it activity alone}. Thus, we find that a moderate contractile activity (e.g., corresponding to emulsions containing actomyosin as the active ingredient) sets up interfacial shear flows which enhance and speed up lamellar ordering. Increasing the strength of contractile activity disrupts the passive lamellar ordering to create emulsions of passive droplets within an active self-stirring background. Instead, we predict that extensile activity (for instance corresponding to mixtures where the active component is a bacterial fluid) should lead to the self-assembly of a polydisperse suspension of active rotating droplets in a passive background. Such a phase may be relevant to the understanding of the self-assembled bacterial rotors reported in~\cite{jana}. By tuning the overall mixture composition, we can also stabilize emulsions in which the majority active phase is dispersed as droplets. This state rapidly undergoes phase inversion as soon as activity is switched off.

Active emulsions like those described here may be self-assembled, by varying component composition, in water-oil emulsions containing actomyosin, or extensile microtubule-motor suspensions~\cite{Dogic}. In the current formulation, the latter active mixture always adsorbs onto the water-oil interface, rather than dispersing inside the aqueous phase~\cite{Dogic}, and the long term stability is also a potential practical problem. However it might be possible to overcome such technical issues in the future -- for instance, hydrodynamic coupling to another structured fluid as in~\cite{Ignes,Ignes2} might be exploited to create active interfacial droplets at the 2D interface. In that case, the surprising richness of behaviour of our exotic active emulsions could be exploited to design active shape-changing microfluidic systems, or to self-assemble soft composite materials with tunable morphology. 

We thank EPSRC for support (Grant EP/J007404/1). Simulations were ran at Bari ReCaS e-Infrastructure funded by MIUR through PON Research and Competitiveness 2007-2013 Call 254 Action I.


\section*{SUPPLEMENTAL MATERIAL}

\section{I. Stress tensor}

The stress tensor $\underline{\underline{\sigma}}^{total}$ considered in the Navier-Stokes equation of the model (see Eq. (2) of the main text)  is the sum of a passive and an active contribution. The former is, in turn, the
sum of three terms, the first of which is the viscous stress $\sigma_{\alpha\beta}^{viscous}=\eta(\partial_{\alpha}v_{\beta}+\partial_{\beta}v_{\alpha})$ ($\eta$ is the shear viscosity)~\cite{note1}.
The second term is the elastic stress and is borrowed from the liquid crystal hydrodynamics. It is written as
\begin{equation}
\sigma_{\alpha\beta}^{elastic}=\frac{1}{2}(P_{\alpha}h_{\beta}-P_{\beta}h_{\alpha})-\frac{\xi}{2}(P_{\alpha}h_{\beta}+P_{\beta}h_{\alpha})
-\kappa\partial_{\alpha}P_{\gamma}\partial_{\beta}P_{\gamma}\label{eq:elastic-stress},
\end{equation}
where $\textbf{P}$ is the polarisation, $\textbf{h}=\delta F/\delta\textbf{P}$ is the molecular field~\cite{deGennes2} and the parameter $\xi$ controls whether the liquid crystal 
molecules are rod-like shaped ($\xi>0$) or disk-like shaped ($\xi<0$). In addition it establishes whether the fluid is flow aligning ($|\xi|>1$) or flow tumbling ($|\xi|<1$) under shear. 
Also, $\kappa$ is the elastic constant of the liquid crystal. The last term includes an interfacial stress and is given by
\begin{eqnarray}
\sigma_{\alpha\beta}^{binary}&=&\left( f-\phi\frac{\delta F}{\delta\phi} \right)\delta_{\alpha\beta} - k\partial_{\alpha}\phi\partial_{\beta}\phi\nonumber\\
 +&c&\left[\partial_{\alpha}\phi\partial_{\beta}(\nabla^2\phi)+ \partial_{\beta}\phi\partial_{\alpha}(\nabla^2\phi)\right]-\beta P_{\beta}\partial_{\alpha}\phi,
\end{eqnarray}
where $F=\int d^3\textbf{r} f$ is the free energy (Eq.(1) of the main text) and $f$ is the free energy density. 
The active contribution is the sole term not stemming from the free energy and is given by~\cite{Simha}
\begin{equation}
\sigma_{\alpha\beta}^{active}=-\zeta \phi \left(P_{\alpha}P_{\beta}-\frac{1}{3}|{\bf P}|^2\delta_{\alpha\beta}\right)\label{eq:active-stress},
\end{equation}
where $\zeta$ is the activity strength that is positive for extensile systems
and negative for contractile ones~\cite{note2}.

\section{II. Lamellar ordering}

\begin{figure*}[htbp]
\centering
\includegraphics[width=17cm]{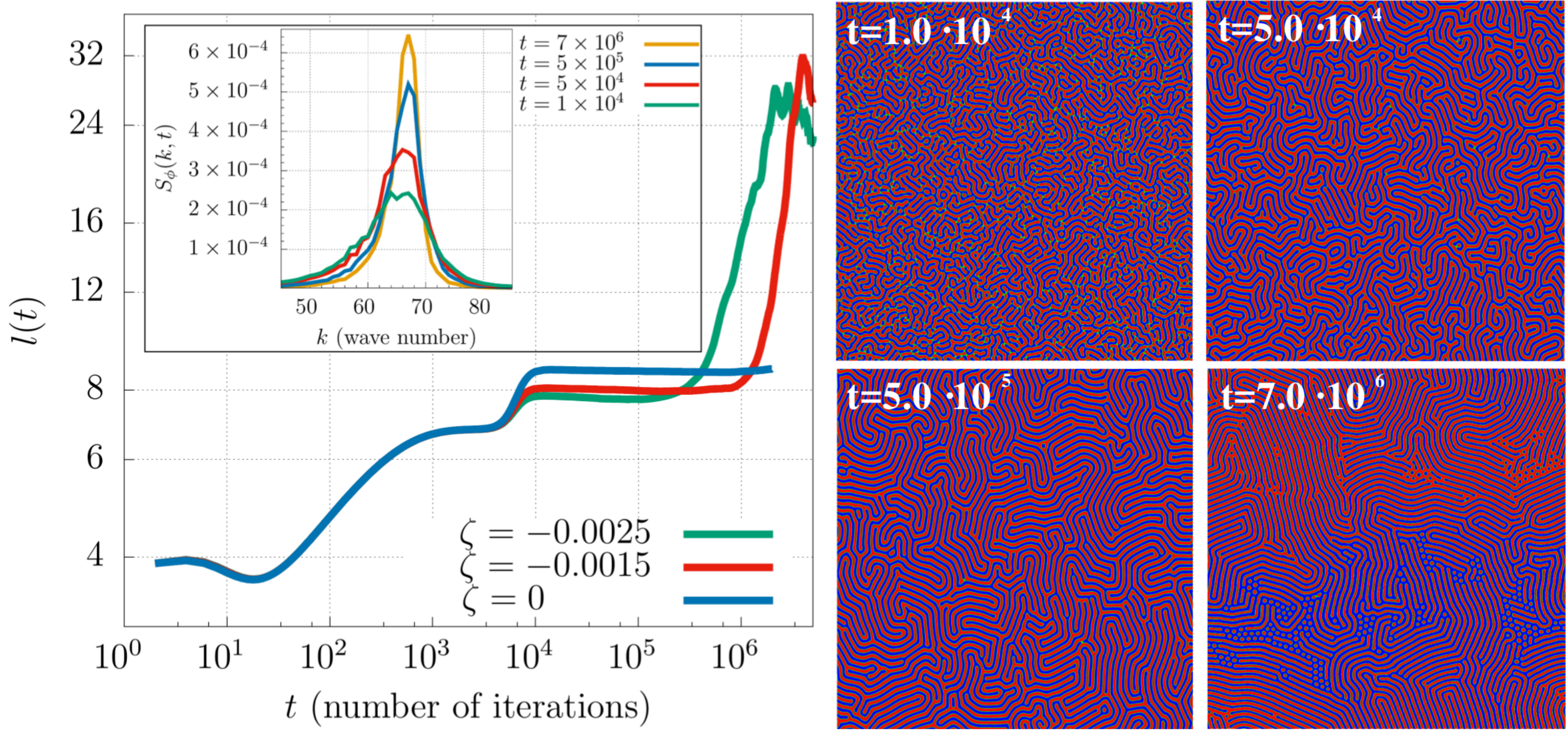}\\[-6pt]
\caption{\textbf{Ordering of lamellar phase at small contractile activity.} (Left) Time-dependent coarsening length $l(t)$ for a 50:50 composition, $\beta=0.01$ and for $\zeta=0$ (blue), $\zeta=-0.0015$ (red) and $\zeta=-0.0025$ (green). While in passive systems phase ordering arrests relatively quickly, higher values of contractile activity speed up the dynamics favouring the formation of more ordered stack of parallel lamellae.
The inset shows the time evolution of $S_{\phi}(k,t)$ for $\zeta=-0.0015$.
After the initial regime, the position of its peak is found roughly at the 
same value of the wave vector $k$, corresponding to 
the lamellar periodicity; the narrowing of its width signals the increase
of order in the system. 
(Right) Time evolution of $\phi$ for $\zeta=-0.0015$ for $\beta=0.01$. Simulations are performed on a square lattice of size $L=512$ and wavevector $k$ in the inset is labeled in lattice units.
The color scale is the same as in the main text, with the active phase in red and the passive phase in blue. This applies to all figures.}
\label{fig7}
\end{figure*}

The ordering dynamics of a $50:50$ contractile emulsion strongly depends on the magnitude of the activity (see Fig.1 of the main text). Indeed, while at early times the domain
morphology observed at different $\zeta$ looks similar as velocities are small and the dynamics is dominated by the diffusive regime, at late times hydrodynamics becomes relevant. If $\zeta$
is small, phase ordering is arrested, whereas for intermediate values
a strong interfacial spontaneous flow drives the formation of almost defect-free regions of parallel lamellae (Fig.~\ref{fig7}, right). A more quantitative
analysis of the lamellar ordering can be achieved by calculating the behaviour of the characteristic length $l(t)$, defined as the inverse of the first moment of the spherically averaged
structure factor  $S_{\phi}(k,t)=\langle\phi({\bf k},t)\phi(-{\bf k},t)\rangle_k$~\cite{kendon} 
\begin{equation}
l(t)=2\pi\frac{\int S_{\phi}(k,t)dk}{\int kS_{\phi}(k,t)dk},
\end{equation}
where $k$ is the modulus of wave vector ${\bf k}$, $\langle\cdot\rangle_k$ is an average over a shell in ${\bf k}$ space at fixed k, and $\phi({\bf k},t)$
is the spatial Fourier transform of $\phi({\bf r},t)$. In Fig.\ref{fig7} (left) we show the time evolution of $l(t)$ calculated on a square lattice of size $L=512$ 
for some values of the contractile activity. Regardless of the value of $\zeta$, the crossover from the early-time diffusion regime to the late one
is always sigmoidal, and lasts up to $t\simeq 10^4$ where a short plateau is observed. Later on ordering restarts, but while for $\zeta=0$ $l(t)$ quickly saturates to a constant
value, for higher values of $\zeta$, after a temporary saturation (the second plateau from $t\simeq 10^4$ to $t\simeq 3\times 10^5$), it rapidly grows as the velocity field becomes strong
enough to disentangle the intertwined domain pattern. However, due to the high number of topological defects in large systems, a full defect-free array of lamellae is difficult to achieve. 
At very late times $l(t)$ may saturate even for the active material, with the formation of few active droplets coexisting with lamellae in a passive polar background. 

\section{III. Morphology of contractile emulsions for $50:50$ composition}

\begin{figure*}[htbp]
\centering
\includegraphics[width = 10cm]{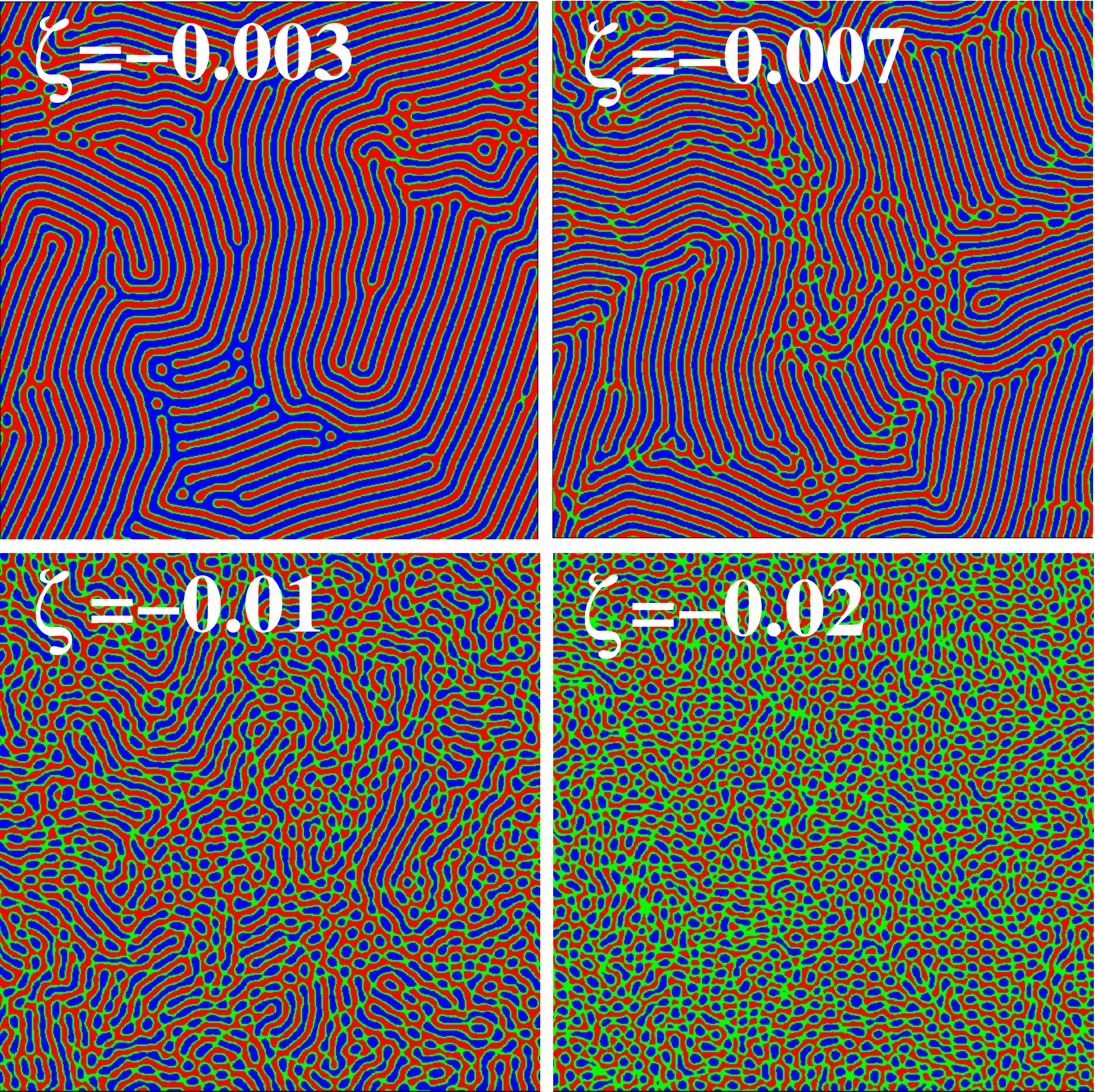}\\[-6pt]
\caption{{\bf Transition from lamellar to emulsion phase at increasing contractile activity.}
We show four snapshots of $\phi$ contour plots for $50:50$ composition 
at late times and for different values of $\zeta$ in a contractile material with $\beta=0.01$. 
While for relatively small values of the activity ($\zeta=-0.003,-0.007$) lamellar ordering dominates, an emulsion of passive material embedded in the active background emerges for higher values ($\zeta=-0.01,-0.02$). Results refer to a square lattice of size $L=256$.}\label{fig6}
\end{figure*}

\begin{figure*}[htbp]
\centering
\includegraphics[width =10cm]{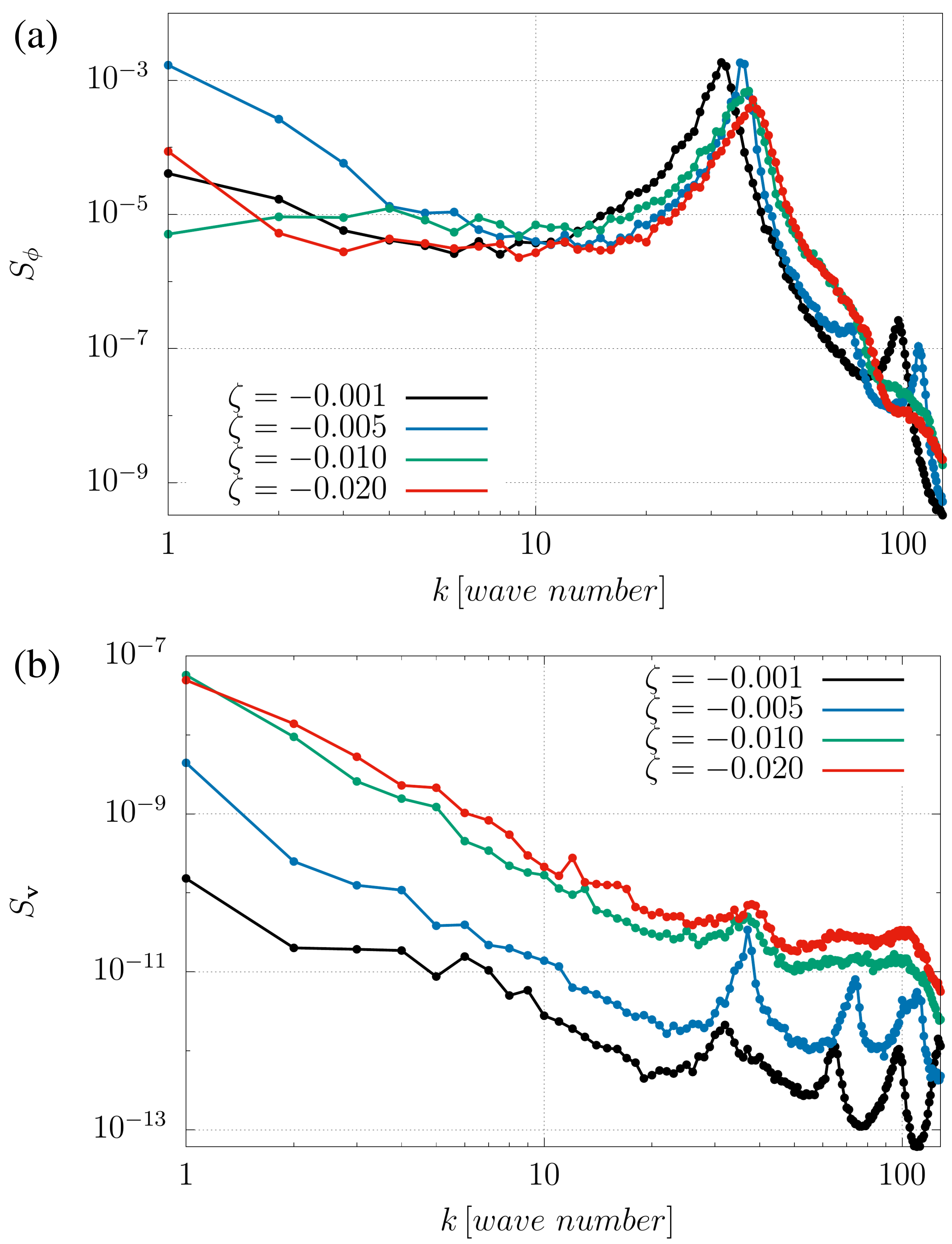}\\[-6pt]
\caption{\textbf{Concentration and velocity structure factor of symmetric contractile mixtures.} (a) Concentration structure factor for different values of $\zeta$ in a contractile material.
An approximately stable peak at  $k\simeq 35$ (corresponding to $l=L/k\simeq 7$) is observed for all $\zeta$ considered. 
A further peak appears at $k\simeq 100$, corresponding to $l\simeq 2.5$, only for $\zeta=-0.001$ and $\zeta=-0.005$.
(b) Velocity structure factor for the same values of $\zeta$. Four peaks appear for $\zeta=-0.001$ and $\zeta=-0.005$
at $k\simeq 38$ ($l\simeq 7$), $k\simeq 80$ ($l\simeq 3.2$), $k\simeq 100$ ($l\simeq 2.5$)  and $k\simeq 105$ ($l\simeq 2.4$), which disappear
for higher values of the activity. Results refer to a square lattice of size $L=256$. Wavevector axis is labeled in lattice units.}\label{fig8}
\end{figure*}

As discussed in the main text, if the activity is sufficiently high, an emulsion of passive droplets assembles in an active contractile background.  
In order to demonstrate how the lamellar-to-droplet transition occurs, we show late-time configurations 
of $\phi$ for $50:50$ composition and for different values of $\zeta$ (Fig.~\ref{fig6}).
While for low values of activity the typical intertwined lamellar pattern dominates (Fig.~\ref{fig6}, $\zeta=-0.003$), 
for increasing values of $\zeta$  small isles of passive material emerge (Fig.~\ref{fig6}, $\zeta=-0.007$).
Further increasing $\zeta$, the mixture undergoes a non-equilibrium transition to an emulsion phase, in which a dynamic array of passive droplets is created in an active matrix  
(Fig.~\ref{fig6}, $\zeta=-0.01,-0.02$ and supplementary movie 4).
A quantitative estimate of the evolution of the characteristic lengthscale $l$ observed by varying $\zeta$ can be obtained by looking at 
the concentration structure factor $S_{\phi}(k,t)$. It shows a peak at $k\simeq 32$ for $\zeta=-0.001$, which shifts towards higher values (e.g. $k\simeq 35$ and $k\simeq 38$) for increasing activity
(Fig.~\ref{fig8}a). They correspond to a lengthscale $l=L/k\simeq 8$, $7$ and $6.7$ respectively, an estimate of the characteristic size either of the lamellae (for low values
of $\zeta$) or of the droplets in the emulsion phase (at high $\zeta$). Note that the steady-state size of the droplets diminishes with increasing contractility.
A further peak appears at $k\simeq 100$ corresponding to $l\simeq 2.5$, an 
estimate of the size of the lamellar interface observed for low values of $\zeta$. For higher values the emulsion
phase dominates and only one characteristic lengthscale appears.
The velocity field structure factor $S_v(k,t)$ (calculated as $S_v(k,t)= \langle{\bf v}({\bf k},t){\bf v}(-{\bf k},t)\rangle_k$), 
besides the two peaks at  $k\simeq 35$ and $k\simeq 100$, displays two further ones at $k\simeq 80$
and $k\simeq 105$ at $\zeta=-0.001$ and $\zeta=-0.005$, corresponding to $l\simeq 3.2$ and $l\simeq 2.4$. 
These indicate the characteristic sizes where intense fluid flows, directed along opposite directions and mainly located at the lamellar interface, are expected to form.
These peaks appear only for small and intermediate values of $\zeta$ whereas they are progressively washed off for high values of $\zeta$.

\begin{figure*}[htbp]
\centering
\includegraphics[width = 10cm]{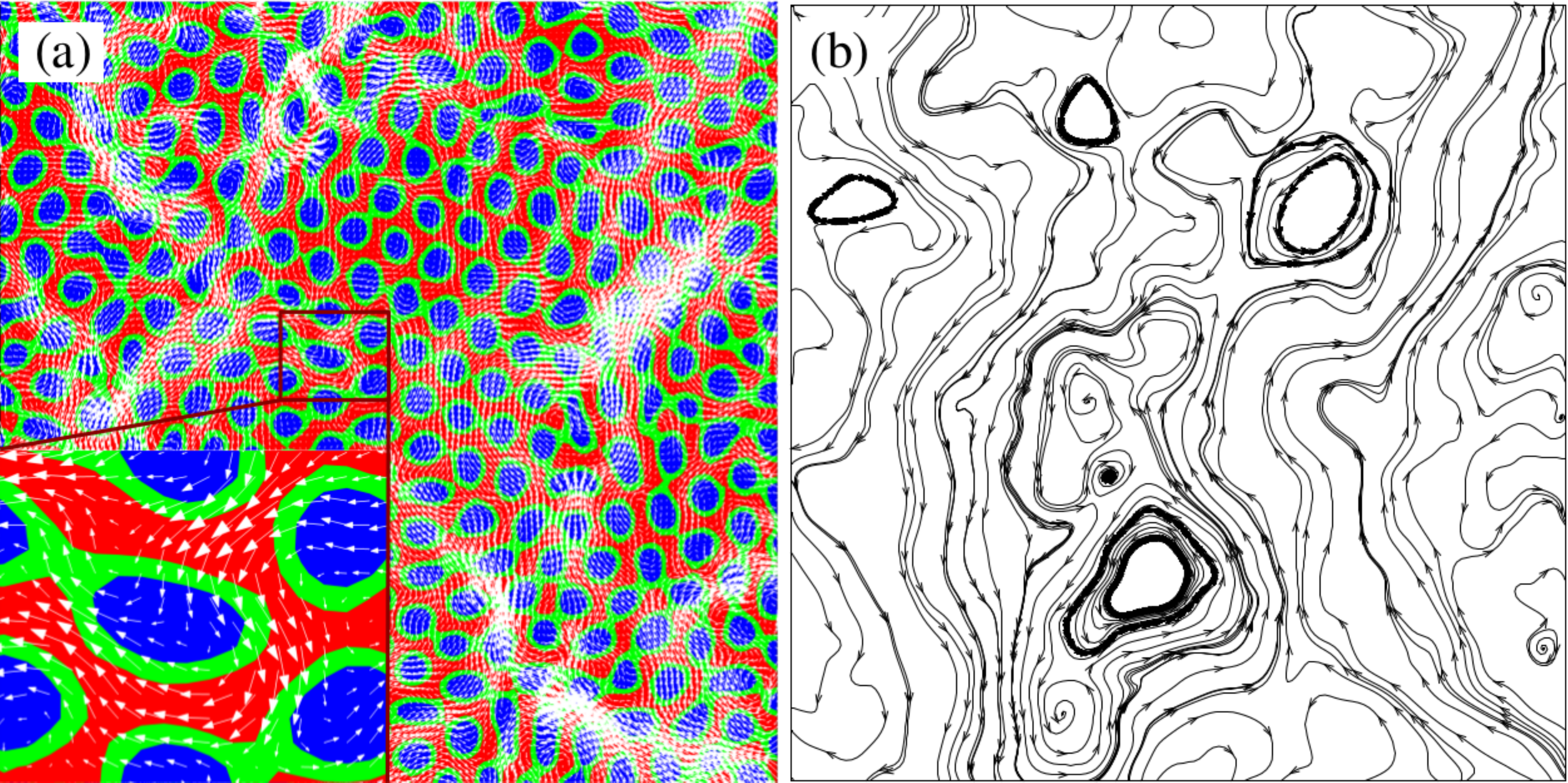}\\[-6pt]
\caption{\textbf{Pattern of the velocity field at high contractile activity}. (Left) Snapshot of the $\phi$ contour plot at late times, with $\beta=0.01$ $\zeta=-0.02$ on a square lattice of size $L=128$. The inset plot shows a zoom of the $\textbf{v}$ field. The spontaneous flow has a typical chaotic signature, which keeps stirring the system and avoids the formation of static structures. 
(Right) Streamlines of the velocity field, whose orientation is defined by the arrows.}
\label{fig1}
\end{figure*}

In order to better understand the structure of the velocity field 
in the emulsion phase, we show it on a square lattice of size $L=128$ (Fig.~\ref{fig1}).
The flow induced by activity keeps stirring the system and favours the formation of rounded droplets. Its chaotic character, not observed for lower values of $\zeta$, 
is shown in the streamlines pattern of Fig.\ref{fig1}b.

\section{IV. Morphology of extensile emulsions for $50:50$ composition}
\begin{figure*}[htbp]
\centering
\includegraphics[width = 10cm]{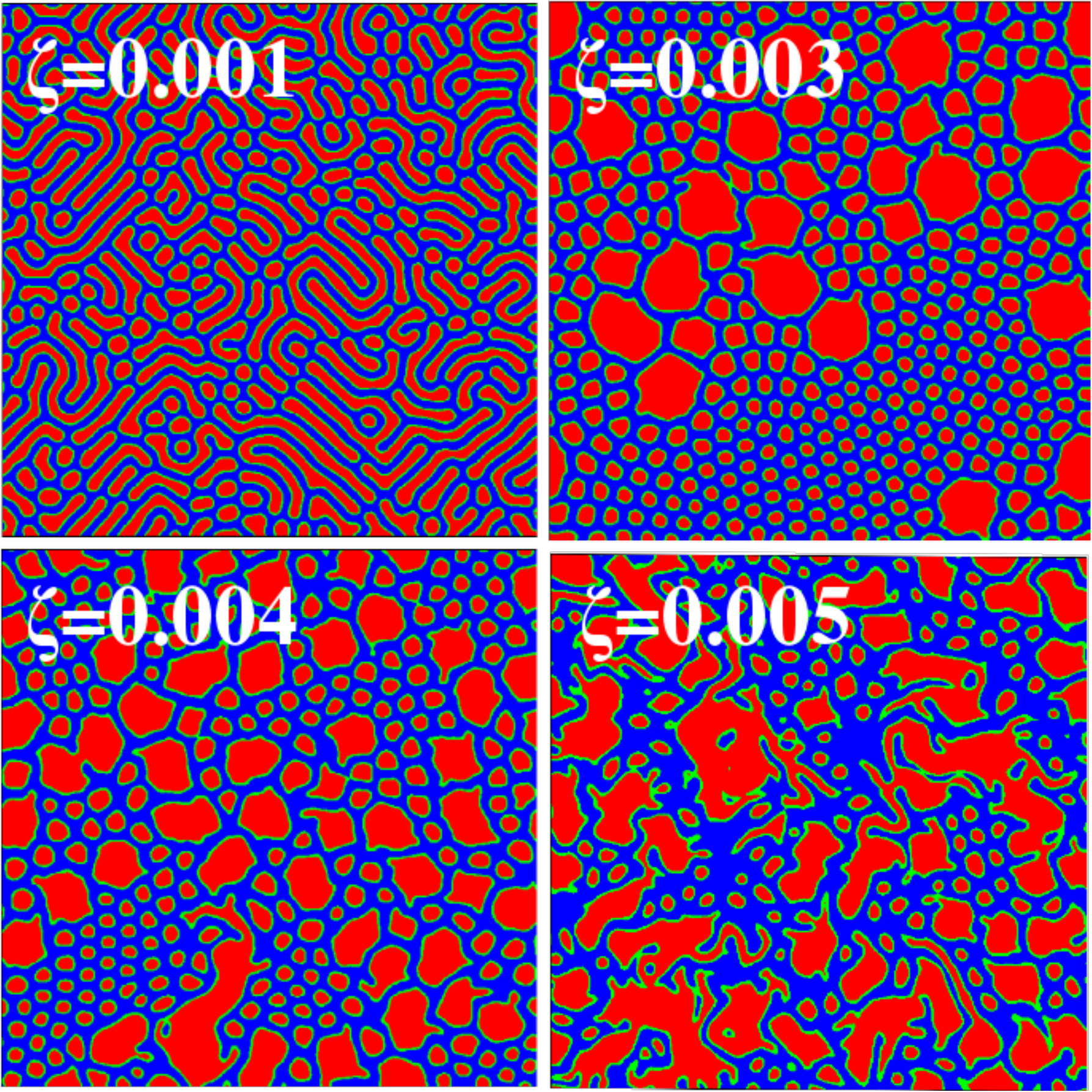}\\[-6pt]
\caption{\textbf{Transition from lamellar to emulsion phase at increasing extensile activity.} Snapshots of $\phi$ contour plots for $50:50$ composition at late times and for different values of $\zeta$ in an extensile material with surface anchoring parameter $\beta=0.01$. 
Here small droplets of extensile material form already at low activity ($\zeta=0.001$). For higher values a bidisperse emulsion emerges ($\zeta=0.003$),
made up by rotating domains whose mean size slightly increases with the activity ($\zeta=0.004$, $\zeta=0.005$) due to collisions and coalescence bewteen them. Results refer to a square lattice of size $L=256$.}\label{fig9}
\end{figure*}

Unlike the case of contractile materials, if the active component is extensile the lamellar phase is lost, and even for small values of $\zeta$ and starting from an initially uniform mixture, an emulsion of active droplets assembles in the passive fluid. In Fig.~\ref{fig9} we show the transition from a structure in which pieces of lamellae can be still observed at $\zeta=0.001$ towards an active emulsion phase in which rounded-shaped domains acquire rotational speed with increasing activity.
\begin{figure*}[htbp]
\centering
\includegraphics[width = 10cm]{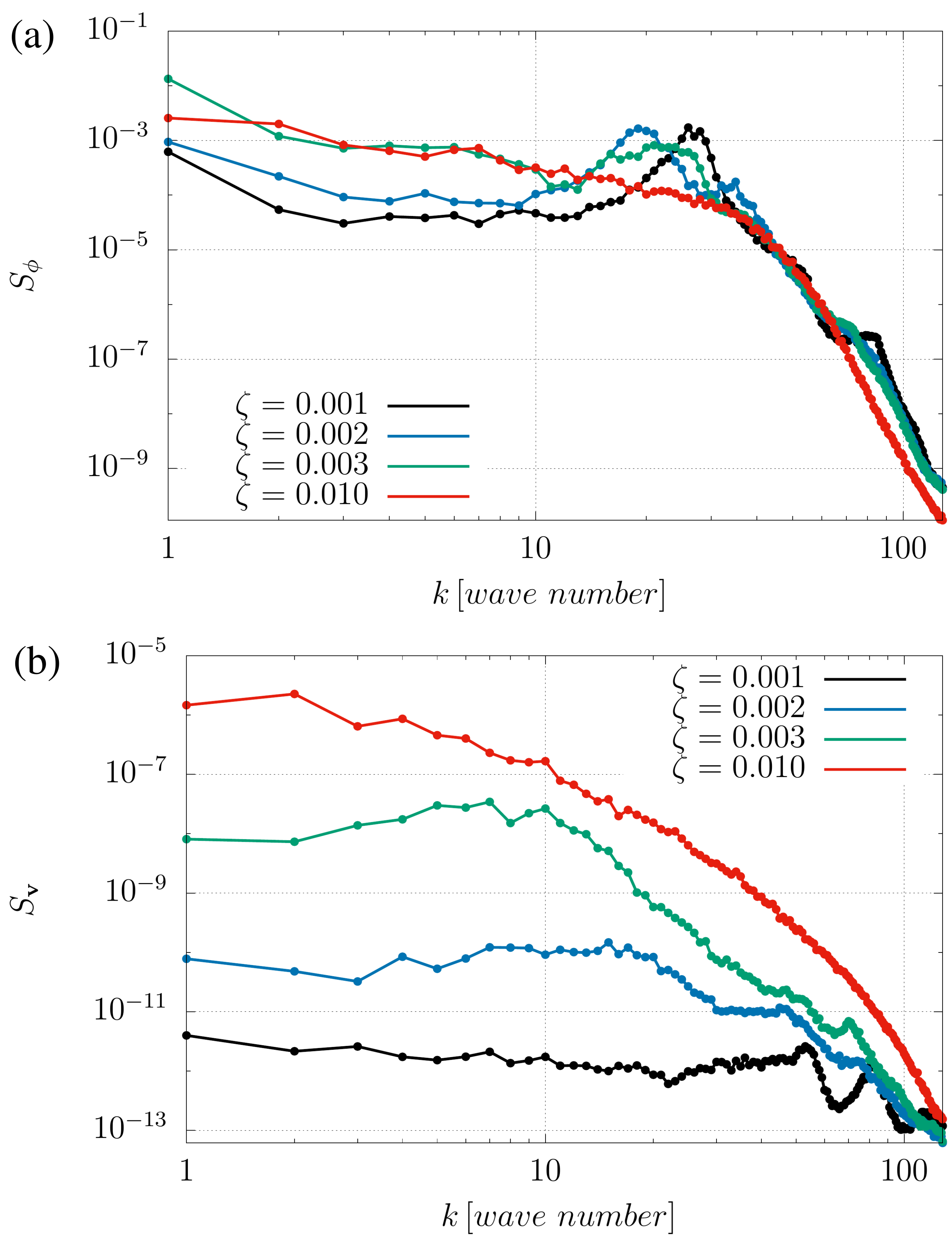}\\[-6pt]
\caption{\textbf{Concentration and velocity structure factor of symmetric extensile mixtures.} (a) Concentration structure factor for different values of $\zeta$ in an extensile material. A broad peak is observed at $k\simeq 20$ corresponds to a lengthscale $l=L/k\simeq 13$ ($L=256$), the characteristic size of the droplets for $\zeta=0.003$.
(b) Velocity structure factor for the same values of $\zeta$. Here $S_v(k,t)$ is roughly smooth for all values of activity.
Wavevector axis is labeled in lattice units.\label{fig10}}
\end{figure*}
The calculation of the concentration structure factor unveils, for instance, a broad peak at $k\simeq 20$ for $\zeta=0.003$, corresponding to $l\simeq 13$, the characteristic lengthscale of the droplets for this case (Fig.~\ref{fig10}a). The velocity structure factor  $S_v(k,t)$ is approximately a smooth curve  in the emulsion phase
(i.e. for values of $\zeta$ higher than $0.002$),  as the fluid velocity inside the droplets and along their interface is overall unidirectional (see Fig.3 of the main text).
It also shows a peak for $\zeta=0.001$ at $k\simeq 90$ corresponding to $l\simeq 2.85$ (Fig.~\ref{fig10}b), the interfacial lamellar distance.
The stronger the activity is the higher the curves are, a further evidence of the increase in the fluid flow strength. Interestingly
each curve in Fig.~\ref{fig10}b exhibits a monotonic decreasing trend, suggesting the typical length-scale over which the velocity field develops.
In particular for higher values of activity length-scales are comparable to the overall size of the lattice; on the contrary the peaks at large wave-numbers for $\zeta=0.001$ demonstrate that, in this case, the dynamical response of the fluid develops over the typical droplet length-scales. 

\section{V. Polarisation pattern and angular velocity in extensile droplets}

\begin{figure*}[htbp]
\centering
\includegraphics[width = 7cm]{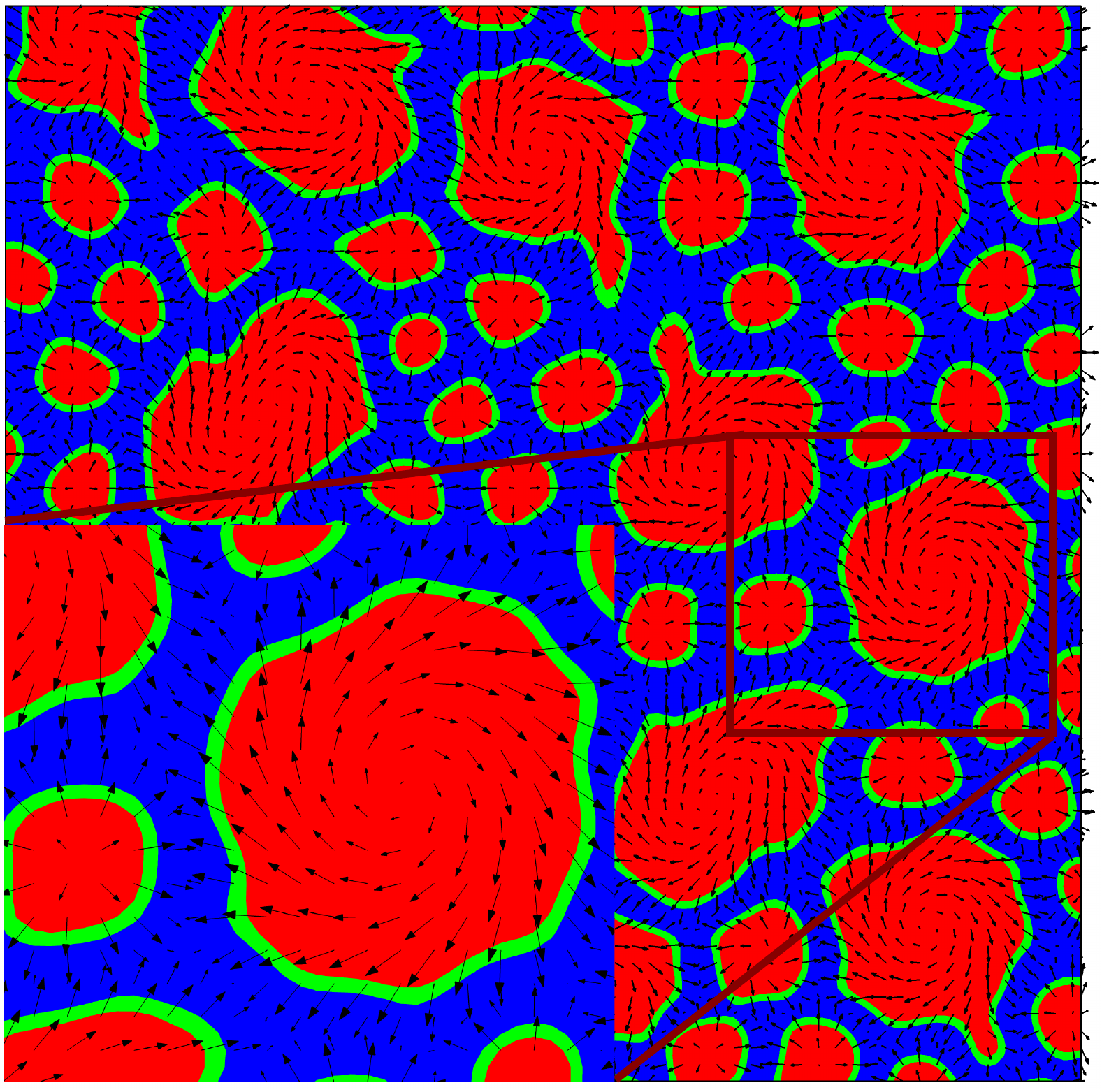}\\[-6pt]
\caption{\textbf{Asters and spiral patterns of the polarization in extensile emulsions.} Snapshots of the $\phi$ contour plot and of the polarization profile at late times, for an extensile material with $\beta=0.01$ and $\zeta=0.004$. Results refer to a square lattice of size $L=128$.}
\label{fig2}
\end{figure*}

\begin{figure*}[htbp]
\centering
\includegraphics[width = 17cm]{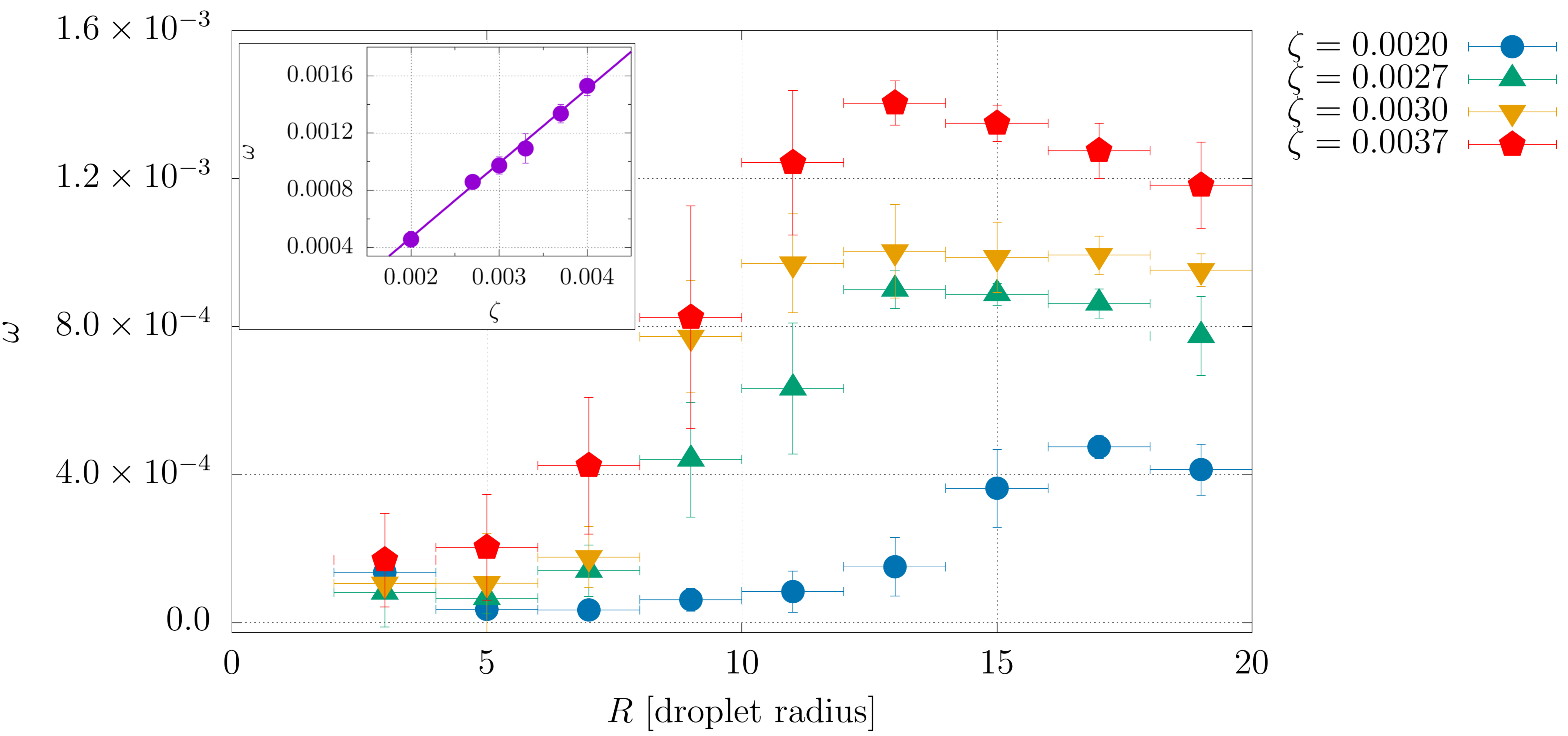}\\[-6pt]
\caption{\textbf{Angular velocity vs. droplet radius.} Angular velocity $\omega$ calculated as a function of the radius $R$ of the droplet for four different values of the activity strength ($\zeta=0.002$  blue/circles, $\zeta=0.0027$ green/triangles, $\zeta=0.003$ yellow/ reverse triangles  and  $\zeta=0.037$ red/pentagons). For all cases the anchoring strength is fixed to $\beta=0.01$. Each value $\omega$ is calculated as an average value over a class of droplets (of at least six uncorrelated configurations at late times) whose radius ranges from $R-1$ to $R+1$ around a defined radius $R$. This also defines the error bar on $x$-axis. The error bar on the $y$-axis is the standard deviation of the average. Note that, at fixed $\zeta$, the angular velocity increases with $R$ up to $R\simeq 13$ for $\zeta=0.0027,0.003,0.0037$, and up to $R\simeq 15$ for $\zeta=0.002$.  Then it remains approximately constant. In this region the mean angular velocity grows linearly with $\zeta$ (see the inset).}\label{fig3}
\end{figure*}

In Fig.\ref{fig2} we show the typical polarisation patterns observed inside an extensile droplet for $\beta=0.01$, $\zeta=0.004$.
While in smaller (non-rotating) droplets the strong perpendicular surface anchoring forces an aster-like pattern in the polarization field,
in the larger ones the high activity yields to the formation of strong bend distortions (the typical elastic instability in extensile materials~\cite{marchetti}), which competes with the surface anchoring. These distortions give rise to spirals in the polarization field and to a vortex-like fluid flow pattern (as that reported in Fig.3 of the main text),
which triggers a spontaneous rotation of extensile droplets when the ratio $\zeta R^2/\kappa$ exceeds a critical threshold,
that we find approximately equal to $13.8$ ($\pm 10\%$). 
A natural question to address is how the angular velocity ${\bf\omega}$ of these droplets depends upon their radius and upon the activity strength $\zeta$. In Fig.\ref{fig3} we report a late time estimate of $\omega$ 
as a function of the droplet radius $R$ for five different values of $\zeta$. 
We calculate the angular velocity as ${\bf\omega} =\frac{\int dV(\phi/R^2) \textbf{r}\times\textbf{v}}{\int dV\phi}$, averaged over a class of droplets whose radius $R$ (the distance from the centre of mass of the droplet) has a bin width of two lattice sites around it.
Clearly $\omega$ increases both with $\zeta$ (as the system possesses higher kinetic energy) and with $R$, up to $R\simeq 13$ for $\zeta=0.0027,0.003,0.0037$, whereas afterwards it remains roughly constant. In this region
dimensional analysis suggests that  $\omega\sim \zeta/\eta$ (with $\eta$ 
effective viscosity); this is also shown in  the inset of Fig.\ref{fig3}, where the mean angular velocity has been plotted versus $\zeta$.
At higher activities the flow field considerably affects
the shape of the droplets, many of which merge into larger domains.

\section{VI. Morphology of extensile emulsions for $10:90$ composition}

\begin{figure*}[htbp]
\centering
\includegraphics[width = 10cm]{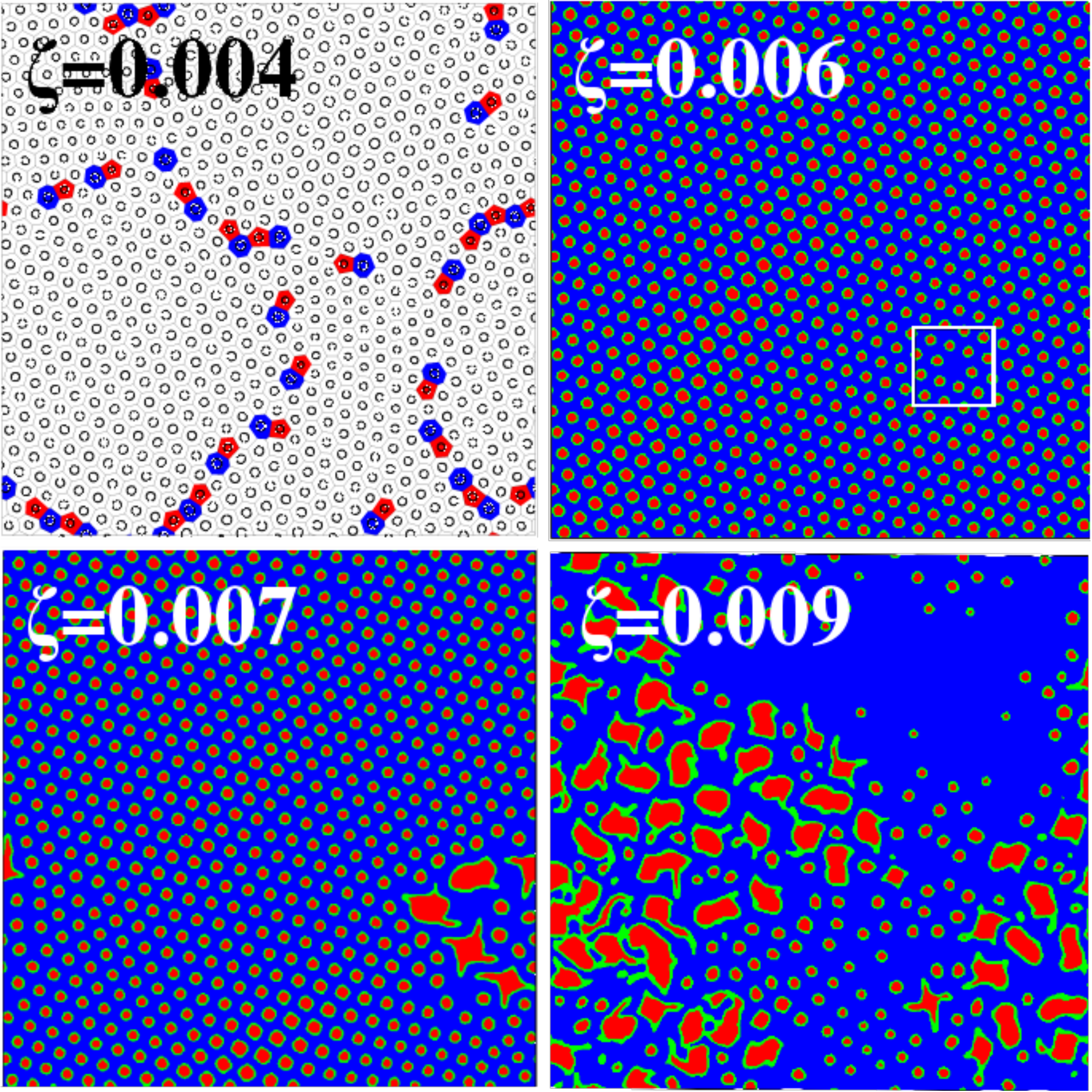}\\[-6pt]
\caption{\textbf{Typical configurations of a highly asymmetric extensile mixture for various $\zeta$.} Snapshots of $\phi$ contour plots with $10:90$ composition, at late times for different values of $\zeta$ in an extensile material with $\beta=0.01$. An emulsion of active droplets forms in an active matrix for all values of $\zeta$ considered. While at low activity ($\zeta=0.004$) droplets arrange in an almost rigid lattice structure with defects, at high activity this arrangement is lost ($\zeta=0.006$, where a vacancy is shown surrounded by a white square) and larger dynamical domains form from collision and merging of smaller droplets ($\zeta=0.007,0.009$). In particular we show the hexatic order observed at $\zeta=0.004$, where droplets with $5$ nearest neighbors are highlighted in red, while those with $7$ neighbors in blue. Results refer to a square lattice of size $L=256$.}\label{fig13}
\end{figure*}
\begin{figure*}[htbp]
\centering
\includegraphics[width = 10cm]{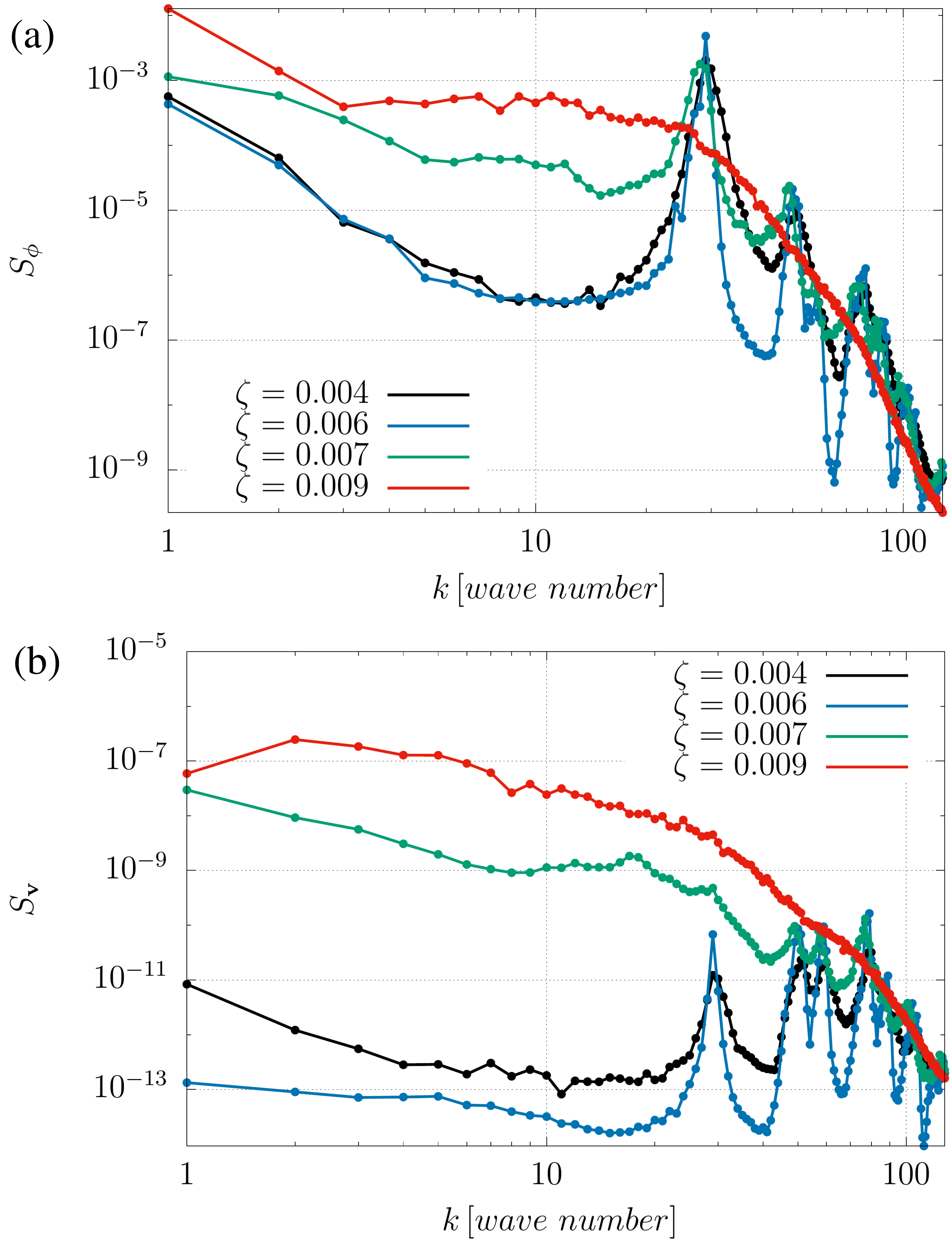}\\[-6pt]
\caption{\textbf{Concentration and velocity structure factor of highly asymmetric extensile mixtures.} (a) Concentration and (b) velocity structure factor for different values of $\zeta$ in an extensile material with $10:90$ composition. Peaks observed at $k\simeq 30$ , $k\simeq 50$  and $k\simeq 80$ correspond to $l\simeq 8.5$, $l\simeq 5.1$ and $l\simeq 3.2$, respectively. The first one gauges the triangular lattice size while the others are two different characteristic lenghts of the domains observed. Wavevector axis is labeled in lattice units. Simulations are perfomed on a square lattice of size $L=256$.}\label{fig14}
\end{figure*}

We have also investigated the dynamics of an extensile emulsion when the area fraction of the active polar gel decreases up to $10\%$. 
Once more we observe a dispersion of active domains (the minority phase) in a passive background (the majority phase), although, unlike the symmetric case, here the active droplets arrange 
in an hexatic order with some defects, as long as $\zeta\lesssim 0.004$ (see Fig.~\ref{fig13}). 
If the activity augments up to $0.006$, a defect-free configuration emerges, except for a vacancy in the droplet lattice (highlighted by a white square in the upper right-hand panel of Fig.~\ref{fig13}).
For more intense active doping ($\zeta \gtrsim 0.007$) the hexatic order is lost and large domains (of radius approximately $R\simeq 10$) of extensile material start to form from the merging 
of several droplets (see the bottom right-hand panel of Fig.~\ref{fig13}). Like the symmetric case they acquire spontaneous rotation, a motion sustained by the combination of strong interface anchoring and intense bend distortions of the polarization field, which, once more, has a spiral structure whose arrows point outwards. Alongside such large domains, smaller non-rotating droplets still appear in the system.
When the activity is very high ($\zeta=0.009$), a higher number of large dynamic domains is created, while the smaller ones progressively shrink by merging/collision and 
probably via Ostwald ripening~\cite{Ostwald}. 
A more quantitative investigation of the behavior of the mixture has been carried out by calculating the concentration and the velocity structure factors (see Fig.~\ref{fig14}). They both show three peaks at $k\simeq 30$, $k\simeq 50$  and $k\simeq 80$, corresponding to $l\simeq 8.5$, $l\simeq 5.1$ and $l\simeq 3.2$, respectively. The first one is a measure of the triangular lattice spacing, on which the centre of an ordered array of droplets will be arranged. The other ones are the characteristic lengths (such as the diameter) of the droplets. Finally note that if the activity is very high ($\zeta=0.009$) the profile is rather smooth and monotonically decreasing over the whole domain in both structure factors. The velocity one, in particular, indicates that fluid develops on the scale of the size of the system.

\section{VII. Phase inversion}
\begin{figure*}[htbp]
\centering
\includegraphics[width = 10cm]{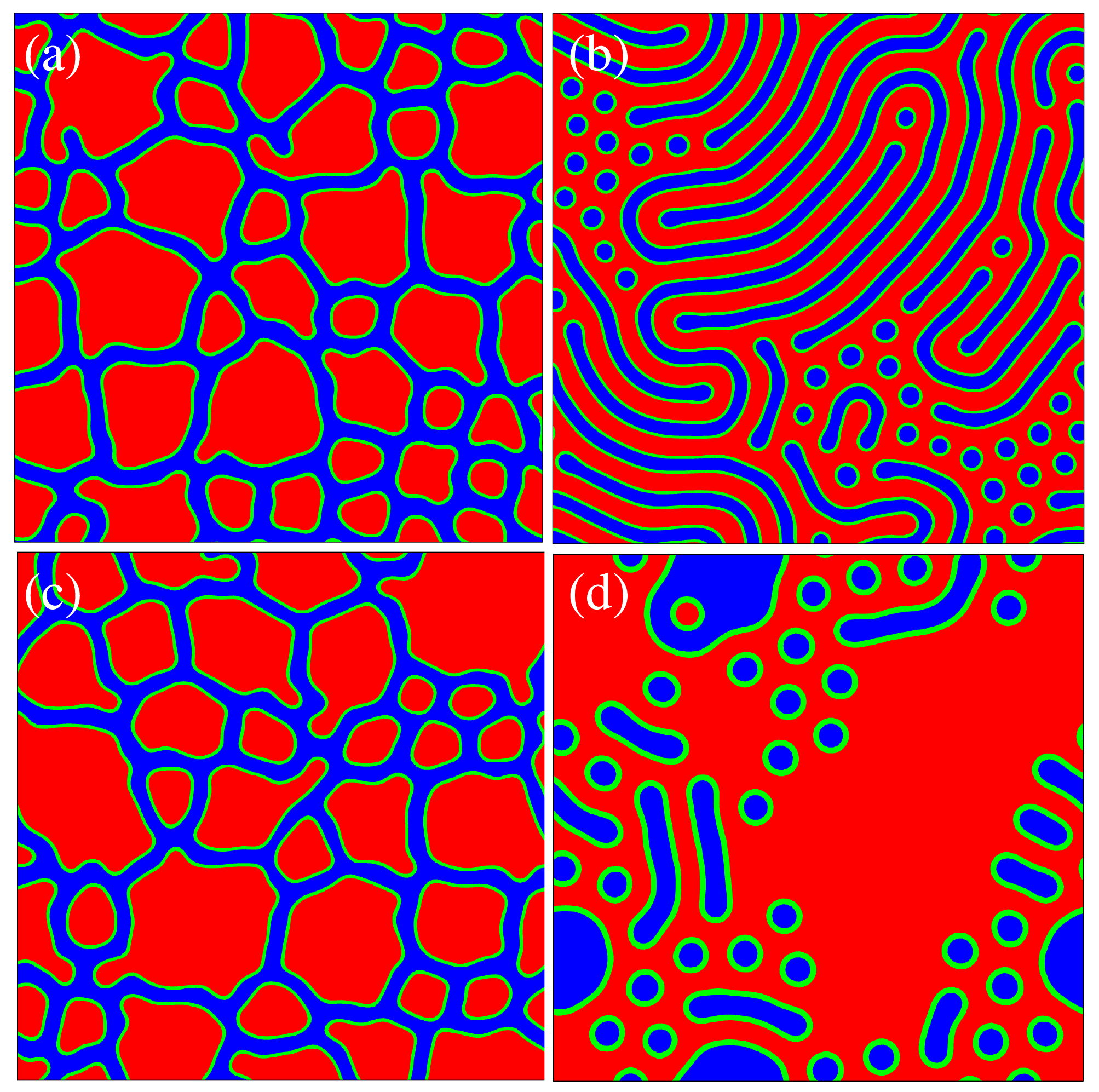}\\[-6pt]
\caption{\textbf{Phase inversion in extensile mixtures.} Panels (a) and (b) show the plot of the concentration $\phi$ at times $t=2\times 10^6$ and $t=3.4\times 10^7$ respectively, for a 70:30 emulsion and $\beta=0.01$. In (a) $\zeta=0.002$, whereas in (b),
$\zeta$ has been switched off to zero. A bicontinuous pattern with isolated isotropic droplets form during time.
Panels (c) and (d) show the plot of the concentration $\phi$ at times $t=2\times 10^6$ and $t=2.2\times 10^7$ respectively, for a 75:25 emulsion. In (c)  $\zeta=0.002$ and $\beta=0.01$ whereas in (d) 
both $\beta$ and $\zeta$ have been switched off to zero. Importantly, here the  phase inversion occurs even when in the absence of surface anchoring.
The switching off dynamics follows the protocol described in the caption of Fig.4 of the main text. Results refer to a square lattice of size $L=128$.}
\label{fig5}
\end{figure*}

In the main text we have shown that if the area fraction of the active gel is 80:20 ($80\%$ of active material and $20\%$ of passive material), emulsions rich of droplets in active material 
are observed. If the activity is switched off, the emulsion undergoes a 
phase inversion, a phenomenon in which the dispersed phase becomes the continuous one and vice-versa. Here  
we corroborate this finding by presenting more results obtained by varying the composition above the critical value 
$0.5$ (see Fig.\ref{fig5}) for a different value of $\zeta$. Isolated regions of active material are again observed
in a passive background. In particular late times configurations of Fig.\ref{fig5}a and c have been obtained for mixtures of respectively 
70:30 and 75:25 compositions, both for $\zeta=0.002$ and $\beta=0.01$. 
Although the two cases look quite similar, if then $\zeta$ is switched to zero 
the resulting dynamics is different. In particular, if the composition is 70:30, the late time  morphology
is a lamellar pattern decorated by small isolated isotropic droplets (see Fig.\ref{fig5}b). If, on the other hand, the composition is 75:25, the late time morphology looks more like a phase inversion 
of isolated islands and droplets dispersed in a continuous polar phase. In this last case $\beta$ has been set equal to zero (namely without surface anchoring). \\

\section{VIII. Numerical details and mapping to physical values}
Eqs.(2-4) of the main text are numerically solved by using a hybrid lattice Boltzmann method (in the limit of incompressible flow), successfully tested in previously studied  systems such as
binary fluids~\cite{binary}, liquid crystals~\cite{liqcrys} and active fluids~\cite{active,active2,active3,active4}. It consists in solving the Navier-Stokes  Equation (2) via a standard lattice Boltzmann 
approach while Eq.(3) and Eq.(4), that govern the time evolution of  respectively 
the concentration $\phi$ and the polarisation $\textbf{P}$ fields, are integrated 
via a finite-difference predictor-corrector algorithm. 
Simulations are performed on a two-dimensional square lattice whose linear 
size ranges from $L=128$ to $L=512$. The system is initialized in a mixed state, with $\phi$ uniformly distributed between $1.1$ and $0.9$ (in symmetric mixtures). 
The concentration $\phi$ ranges from $\phi=0$ (passive phase) to $\phi\simeq 2$ (active phase), that correspond to  the two minima of the double-well potential. 
The starting polarization field $\textbf{P}$ is randomly distributed 
between 0 (passive phase) and 1 (active phase). 

By following previous studies~\cite{active2,active4}, an approximate relation 
between simulation units and physical ones (such as those of a contractile active gel) can be obtained by using
as length-scale, time-scale and force-scale respectively the values $L=1\mu$m, $\tau=10$ms and $F=1000$nN (see Table.~\ref{table1}). 
Note that, as in previous Lattice Boltzmann simulations, the fluid mass density $\rho$ is much larger than the mass density of a real solvent (such as water)~\cite{cates}. 
This assumption that is valid as long as inertial effects are negligible compared to viscous ones, 
reduces the computation time by several orders of magnitude. 
Throughout our simulations the Reynolds number, for the case in which the droplets are observed, is evaluated in terms of the average droplet radius, of the viscosity and of the velocity of the fluid.
It remains below $0.1$, a value in which inertial effects are indeed negligible.

\begin{table}[htbp]
\caption{Typical values of the physical quantities used in the simulations.}
\label{table1}
\vskip 0.3cm

\begin{tabular}{p{3.8cm}|c|c}
Model variables and parameters & Simulation units & Physical units  \\
\hline
Effective shear viscosity, $\eta$                        & $5/3$                 & $1.67\, \mathrm{kPa}\mathrm{s}$ \\
Effective elastic constant, $\kappa$                     & $0.006$               & $6\, \mathrm{nN}$ \\
Shape factor, $\xi$                                      & $1.1$                 & dimensionless \\
Effective diffusion constant, $D=Ma$                     & $0.0004$              & $0.004\,\mu \mathrm{m}^{2}\mathrm{s}^{-1}$ \\ 
Rotational viscosity, $\Gamma$                           & $1$                   & $10\, \mathrm{kPa}\mathrm{s}$ \\
Activity, $\zeta$                                        & $0-0.01$              & $(0-100)\, \mathrm{kPa}$ \\
\end{tabular}
\end{table}

\section{IX. Supplemental movies}
{\bf Supplementary Movie 1:} This movie shows the dynamics of a contractile  active material (with a $50:50$ ratio of active and passive components)  when $\beta=0.01$ and $\zeta=-0.002$
(Fig.1(a) of the main text)

{\bf Supplementary Movie 2:} This movie shows the dynamics of a contractile  active material (with a $50:50$ ratio of active and passive components)  when $\beta=0.01$ and $\zeta=-0.006$
(Fig.1(b) of the main text)

{\bf Supplementary Movie 3:} This movie shows the dynamics of a contractile  active material (with a $50:50$ ratio of active and passive components)  when $\beta=0.0$ and $\zeta=-0.006$
(Fig.1(c) of the main text)

{\bf Supplementary Movie 4:} This movie shows the dynamics of a contractile  active material (with a $50:50$ ratio of active and passive components)  when $\beta=0.01$ and $\zeta=-0.02$
(Fig.2 of the main text and Fig.\ref{fig1} of the SI)

{\bf Supplementary Movie 5:} This movie shows the dynamics of an extensile  active material (with a $50:50$ ratio of active and passive components)  when $\beta=0.01$ and $\zeta=0.002$
(Fig.3 of the main text)

{\bf Supplementary Movie 6:} This movie shows the dynamics of an extensile active material (with a $50:50$ ratio of active and passive components) when $\beta=0.01$ and $\zeta=0.003$ (Fig.3 of the 
main text and Fig.\ref{fig2} of the SI).

\if[

\fi


\begin{thebibliography}{99}
\bibitem{note2} In the derivation of the active stress tensor higher order terms (such as $\partial_{\alpha}P_{\beta}$) are allowed by symmetry. Our assumption is anyway
acceptable as the degree of asymmetry of elongated fibres (e.g. actin filaments) is supposed to be small.
\end{thebibliography}
\end{document}